# Affective Automotive User Interfaces – Reviewing the State of Driver Affect Research & Emotion Regulation in the Car

MICHAEL BRAUN, BMW Group Research, New Technologies, Innovations and LMU Munich, Germany
FLORIAN WEBER, BMW Group Research, New Technologies, Innovations, Germany
FLORIAN ALT, Bundeswehr University and LMU Munich, Germany

Affective technology offers exciting opportunities to improve road safety by catering to human emotions. Modern car interiors enable the contactless detection of user states, paving the way for a systematic promotion of safe driver behavior through emotion regulation. We review the current literature regarding the impact of emotions on driver behavior and analyze the state of emotion regulation approaches in the car. We summarize challenges for affective interaction in form of technological hurdles and methodological considerations, as well as opportunities to improve road safety by reinstating drivers into an emotionally balanced state. The purpose of this review is to outline the community's combined knowledge for interested researchers, to provide a focussed introduction for practitioners, raise awareness for cultural aspects, and to identify future directions for affective interaction in the car.

CCS Concepts: • **Human-centered computing** → **Human computer interaction (HCI)**.

Additional Key Words and Phrases: Affective Computing, Automotive User Interfaces, Emotion Regulation

## 1 INTRODUCTION

Advancing research in the fields of ubiquitous sensing and machine learning now allow for user-awareness in digital products, such as voice assistants or automotive user interfaces. Human-computer interaction can benefit from understanding the user, as it allows for an adaptation to their current state of attention, health, emotions, etc. – and thus enables a more natural way of interaction. Within the research field of *Affective Computing*, we envision systems which can, for example, detect the emotional state of users and accordingly act in an empathic way. The word *affective* connotes "relating to, arising from, or influencing feelings or emotions" [26]. Affective user interfaces stand out as they "sense, interpret, adapt, and potentially respond appropriately to human emotions" [67]. Since the initiation of this research area in the late 20th century by Rosalind Picard [83], sensing emotions through psycho-physiological measures, speech analysis, or facial expressions has been one of the main focus points for researchers in the field. Applications for the technology have naturally been thought of and over time have taken shape along the progress of sensing approaches. Among them, affective systems for in-car usage are seen as a possibly beneficial implementation, as emotions can have serious impacts on road safety [31].

Authors' addresses: Michael Braun, BMW Group Research, New Technologies, Innovations;, LMU Munich, Munich, Germany, michael.bf.braun@bmw.de; Florian Weber, BMW Group Research, New Technologies, Innovations, Munich, Germany, florian.ww.weber@bmw.de; Florian Alt, Bundeswehr University;, LMU Munich, Munich, Germany, florian.alt@unibw.de.

A decade has passed since Eyben et al. provided a literature review on affective technologies which could be used to improve driving safety and the user experience in the car [31]. In the meantime, the research community investigating the connection between in-car HMI, driving safety, user experience and the driver's emotional state has evolved. Apart from studies on the effects of emotions on driving and the context of emotions on the road, also interaction approaches with the intent to regulate the driver's emotional state have been evaluated. We are now at a point in the development of affective automotive user interfaces where research is taking influence on mass produced vehicles and thus product designers, engineers, and researchers need a common understanding of the state of the art. This literature survey summarizes the efforts to understand the principles of emotions on the road, the effects of emotional states on driving safety, the contexts they evolve in, and how user interfaces can regulate detected driver states.

We conduct a systematic review within the digital libraries of the most relevant publishers in HCI (ACM, Elsevier, IEE, Springer) and an additional search on Google Scholar to identify relevant publications released elsewhere. The basic body was received using the following search query: *[affective || emotion-aware || emotion || emotions || emotional] && [automotive || car || cars || in-car]*, resulting in a total of 3633 matches. An ensuing classification regarding topic fit, as many results dealt with, e.g., affect disorders or image processing algorithms, which are not at the focus of this review, delivered a set of 131 relevant publications. We then eliminated duplicates, as the used sources are not mutually exclusive and performed a snowball search of the referenced work within the identified set of publications. This provided a retrospective set of additional publications which either used different wordings or were published in less accessible formats and thus escaped the initial search. We did not include unpublished works in our review. Following this collection phase, we analyzed the manuscripts regarding their research goals and the methods incorporated and divided them into several topic sections to obtain a clear structure:

**Section 2** provides reading pointers for emotion detection, as it is not the focus of this survey.

**Section 3** describes the effects of emotions on driver behavior, laid out in detail in Table 1.

**Section 4** covers research on triggers of emotions while driving, visualized in Table 2.

**Section 5** aggregates experimental evaluations of emotion regulation approaches, see Table 3.

**Section 6** summarizes recent industry applications and shows related research ideas in Table 4.

**Section 7** discusses the methods used in the given research, with an overview in Table 5.

## 2 EMOTION DETECTION IN THE CAR

Emotions are broadly characterized as human responses accompanied by distinct patterns of conscious or unconscious psycho-physiological activity [28], allowing for individual variability as they are contingent on appraisal [40]. They are often categorized in six basic emotions [29] or organized in two continuous dimensions of valence and arousal [90].

Emotion detection techniques are not at the focus of this work. Thus, we only provide a limited selection of surveys and research cornerstones we recommend as introductory readings into the topic. Recent emotion detection algorithms build upon the psychological frameworks of Russell's circumplex model of emotions [90] and the basic emotions introduced by Ekman et al. [29]. Thanapattheerakul et al. give an overview on preceding emotion theories from Darwin to Russell and the current status of research [101]. They also explain how neuroimaging, implemented through functional magnetic resonance imaging (fMRI) or electroencephalogram (EEG) and reactions of the autonomic nervous system (ANS) can be used to detect emotions. As an introduction from a psychological point of view, we recommend the work of Feldman Barrett et al., pointing out the challenges of emotion detection with current techniques [9] and the summary of emotion detection efforts, based on EEG by Thirunavukkarasu et al. [102]. Vogt et al. further provide an early literature review focussed on detecting emotions from speech [109] and Poria et al. expand on this with emotion detection approaches from auditory data (automatic speech recognition, vocal affect) and visual inputs (body gestures, facial expressions) and multimodal combinations thereof [84].

The fact that emotional reactions also influence body temperature, especially in areas with many surfacing blood vessels like facial regions, can also be exploited to detect affect with thermal imaging technology [1]. Related work also compares commercial emotion detection applications using speech, facial expressions, written text, body gestures or movements and physiological states, arriving at the conclusion that sensor fusion is the best approach if possible [38]. Two recent dissertations show that especially for use in the car, unobtrusive solutions with implicit input streams are favored to detect emotions: the work of Marlene Weber is based on the analysis of facial expressions as a contactless application [114], while Jennifer Healey focussed on wearables for affect recognition from physiology [46]. Driving behavior like acceleration and steering angles [96] as well as more obtrusive sensing methods like EEG headbands [45] and physiological sensors [32, 93] have likewise been applied to successfully detect affective states while driving.

Although highly sophisticated technology has been deployed to solve the problem of emotion detection, current approaches are all but undisputed within the science world. As an example, Feldmann-Barrett et al. criticize that available evidence from different populations demonstrates facial expressions of emotions are not universal to humankind but heavily depend on context [9]. Conn Welch et al. further challenge the practicality and compatibility aspects of emotion recognition technology in automotive contexts [115]. Current applications are also incapable of detecting subtle or complex states, and cannot cope well with with noisy sensor data. Research on affective interfaces thus often works with induced emotions or wizard approaches, assuming more matured detection in the foreseeable future.

## 3 EFFECTS OF EMOTIONAL STATES ON DRIVING

The primary motivation for affective automotive user interfaces comes from the fact that the emotional state of the driver inevitably influences their behavior and thus has an effect on road safety. One aspect of relevant behavioral principles was postulated in 1908 as the Yerkes-Dodson law, stating that task performance is interconnected with arousal, specifically that performance is best with medium levels and worst with high or low levels of arousal [118]. Multiple researchers have applied this concept to the driving context and extended the initial measure of arousal with a second dimension of valence, following Russell's theory of a two-dimensional affect model [90]. The common understanding among traffic researchers is that driving performance is best in medium arousal and medium to high valence. Figure 1 shows a surface plot modeled after work by Cai & Lin [21]. They criticize that high valence usually induces increased arousal, which in turn influences performance negatively. This inspires the approach followed by many researchers to assume a safe driving state when the driver has medium levels of arousal and generally positive but not necessarily a maximum of valence. Figure 2 illustrates the expected relation in Russell's two dimensions.

Related work often rather works with categorical than dimensional emotion classifications, with anger being the by far most investigated driver state. These states are easier to induce and detect than continuous valence-arousal levels, and thus more realistic in applied research [55]. According to Jeon & Walker, the most relevant emotions due to their negative effects on road safety are Anger, Fear and Happiness, which accounted for 48% of variance in a qualitative study with 14 participants [54]. Further basic affective states dealt with in literature are Sadness and Contempt. Roidl et al. additionally indicate the importance of Surprise [88], which however is left as the only basic emotion so far not addressed in automotive research. As surprise generally entails increased mental workload [86], we assume that traffic researchers did so far not see a benefit in investigating this specific emotional state, as the impacts of surprised behavior might be covered in research on distraction and cognitive load, e.g. by Engström et al. [30].

The existing literature on the effects of emotions on driving is set out in Table 1, partitioned into categorical and dimensional emotions, as well as several impact areas. The research distinguishes between effects of emotions on aggressive driving, driving performance, reaction time, risk perception, risk taking, road element localization, subjective safety level and workload. Results have only been included if they were achieved using experimental research settings, e.g. in lab or simulator studies, or as a result of the analysis of naturalistic driving data, and if a baseline state, e.g. neutral emotion, was induced or measured in order to ensure comparable results. Some emotions, like surprise, are seldom researched in the automotive domain because of their inherently distracting nature. We only included affect categories for which we actually found published research articles.

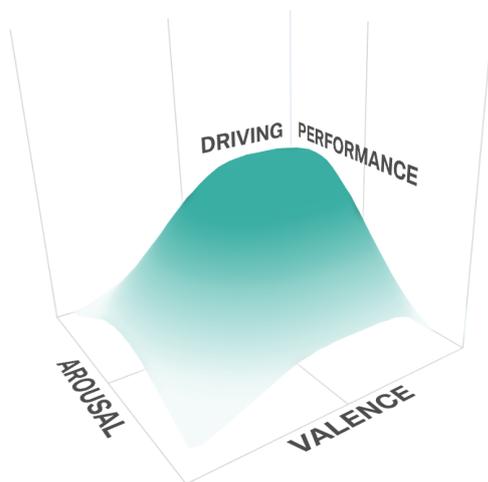
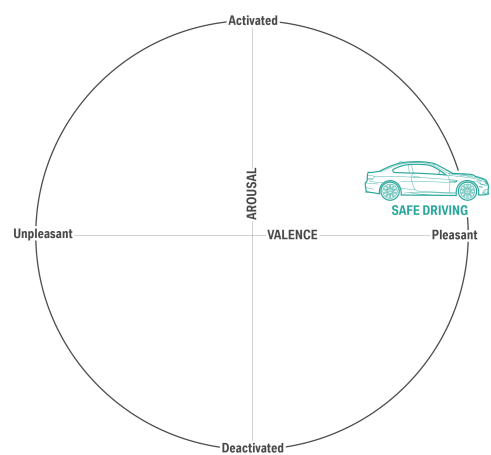

Fig. 1. Model of driving performance based on valence and arousal as proposed by Cai & Lin [21]. Best performance is expected at medium arousal, derived from the law of Yerkes & Dodson [118].

Fig. 2. Emotional driver taxonomy on grounds of the circumplex model by Russell [90]. Assumptions of a safe driving state at medium arousal and high valence predominate in the literature [12].

### 3.1 Aggressive Driving

Under the terms aggressive and reckless driving we understand deliberate unsafe driving behaviors, as opposed to unintentional errors affecting driving performance [85] with often "intentional acts of bodily and/or psychological aggression toward other drivers" [27]. Aggressive driving is hypothesized to be connected directly with the emotional state of Anger: Findings from an on-road study with 44 participants by Mesken et al. [70], as well as the analysis of driving diaries of 100 British drivers by Underwood et al. [106] and data from the SHRP 2 naturalistic driving study analyzed by Precht at al. [85] support this theory. Evidence points towards a direct relationship between a driver state of Anger and aggressive driving, shown in driving simulator experiments with 24 Chinese drivers conducted by Zhang et al. [120] and 40 UK drivers by Zimasa et al. [122]. A series of 5 experiments with a total of 571 Israeli students by Taubman - Ben-Ari also suggest a greater willingness to drive recklessly in positive emotional states, although only when combined with high arousal [100].

# Effects of Emotional States on Driving

| | Anger | Fear | Happiness | Sadness | Contempt | Valence | Arousal |
|---|---|---|---|---|---|---|---|
| **Aggressive Driving** | Mesken et al. 2007<br>Precht et al. 2016<br>↑ Underwood et al. 1999<br>Zhang et al. 2016<br>Zimasa et al. 2019 | | | | | | ↑ Taubman - Ben-Ari 2011 |
| **Driving Performance** | ⊘ Precht et al. 2016<br>Deffenbacher et al. 2003<br>Jeon et al. 2011, 2014, 2016<br>↓ Roidl et al. 2014<br>Steinhauser et al. 2018<br>Zhang et al. 2016 | Jeon et al. 2014<br>↓ Roidl et al. 2014 | ↓ Jeon et al. 2014 | ↓ Jeon 2016 | ↓ Roidl et al. 2014 | ∩ Cai & Lin 2011<br>↑ Chan & Singhal 2014<br>Trick et al. 2011 | ∩ Cai & Lin 2011 |
| **Reaction Time** | Jallais 2014<br>↑ Steinhauser et al. 2018<br>Zimasa et al. 2019 | | ↑ Steinhauser et al. 2018 | | | ↑ Trick et al. 2011 | ↑ Trick et al. 2011 |
| **Risk Perception** | ⊘ Jeon 2016<br>Lee 2010<br>↓ Lu et al. 2013 | ↑ Lu et al. 2013<br>Mesken et al. 2007 | | ⊘ Jeon 2016 | | ↓ Hu et al. 2012 | |
| **Risk Taking** | ↑ Hu et al. 2012<br>Li et al. 2019 | | | ↑ Hu et al. 2012 | | | |
| **Road Element Localization** | ↓ Zhang et al. 2016 | | | ↓ Jallais et al. 2014 | | | |
| **Subjective Safety Level** | ↓ Jeon et al. 2014 | ⊘ Jeon et al. 2014 | | | | | |
| **Workload** | ⊘ Jeon et al. 2014<br>↑ Jeon 2016 | ⊘ Jeon et al. 2014<br>↑ Jeon et al. 2011 | ⊘ Jeon et al. 2014 | | | | |

**Legend**
↑ Positive connection
↓ Negative connection
⊘ Comparable to neutral
∩ Downward parabola

Table 1. Effects of affective driver states on safety relevant driver behavior, grouped by categorical emotions, valence and arousal.

## 3.2 Driving Performance

Driving performance is defined by the Society of Automotive Engineers as measures of longitudinal (e.g., distance headway, time-to-collision) and lateral control (e.g. standard deviation of lane position, lane departures) [51]. Many researchers have investigated the effects of Anger on driving performance, with most of them concluding that Anger has a negative effect on driving performance. A set of experiments with US undergraduate students in low-fidelity driving simulations by Deffenbacher et al. (121 participants) [25] and Jeon et al. (24/70/61 participants) [53, 55, 56] gather evidence for this connection, as well as studies in static high-fidelity driving simulators with 79 and 90 German drivers by Roidl et al. [87] and Steinhauser et al. [99] and 24 Chinese drivers by Zhang et al. [120]. Only Precht et al. report from their analysis of SHRP 2 naturalistic driving data that angry drivers show comparable driving performance as drivers in a neutral state [85]. They distinguish the effects of Anger on driving performance and aggressive driving, meaning that angry drivers are not inhibited due to decreased performance but actively chose to drive aggressively.

In addition to Anger, the studies presented above have yielded disparate findings for Fear: Roidl et al. report detrimental effects on driving performance [87], while Jeon et al. find no evidence for such effects [55]. Further reported findings also give account of negative effects on driving performance resulting from extensive Happiness [55], Sadness [53] and Contempt [87].

Cai & Lin abstract the essence of these relations into a model of parabolic influences of arousal and valence (see Figure 1), meaning high and low levels decrease driving quality compared to neutral levels, which they consolidate in a low-fidelity simulator study with 15 post-/graduate students. They state that high valence might not have a negative impact on its own but usually comes accompanied by high arousal, which again deteriorates driving performance. Two driving simulator studies with 25 and 26 North American university affiliates by Chan & Singhal [23] and Trick et al. [105] confirm this positive relationship between driving performance and valence levels.

## 3.3 Reaction Time

The effect of emotions on reaction time as a sub-measure of driving performance has been investigated by a series of studies. Reaction time in a driving context is defined by the SAE as the time "from onset of an initiating event to the first movement of the driver's hand (on the steering wheel) or foot (on a pedal or from the floor)" [51]. A lab study conducted by Jallais et al. with 54 French drivers suggests increased reaction time for angry participants, a finding which is similarly encountered in the work of Zimasa et al. [122] and Steinhauser et al. [99], whereas the latter also report that Happiness has a comparable effect. Trick et al. report that both increased valence and arousal are affecting reaction time in a negative way [105].

## 3.4 Risk Perception

Several studies investigated the effects of emotions on risk perception, meaning the intuitive judgement of traffic situations and possible consequences [97]. Jeon performed an experiment in simulated driving conditions, where 61 college students showed no significant differences in risk perception for the emotional states Anger and Sadness. They find this especially alarming since these states are assumed to deteriorate driving performance, which might in combination amplify the disposition for hazardous maneuvers [53]. The work of Lee gives account of a driving simulator study with 24 US drivers, whose risk perception capabilities were negatively influenced by induced anger [62]. Lu et al. also report a reduction in risk perception for angry drivers in a lab study with 97 Chinese participants [66]. Furthermore, Fear [66] and Anxiety [70] have been identified as affecting subjective risk judgement positively. Hu et al. report a general negative correlation of risk perception and valence levels from an online survey with 500 and an ensuing lab study with 218

Chinese participants [50]. Jeon pinpoints this interplay between risky behavior and the awareness thereof as crucial for driving safety, stating that highly aroused (e.g., angry) drivers are usually aware of the fact that they are endangering themselves or others, while drivers in a negative low arousal state (e.g., Sadness) showed a lack of self-perception and consequently did not compensate for their dangerous driving [53].

### 3.5 Risk Taking

Risk taking is distinguished from aggressive driving as potentially dangerous driving behavior without aggressive intentions toward oneself or other traffic participants [27]. It is suggested to be influenced by the emotional state of Anger, as Li et al. found in a lab study with 24 Chinese university affiliates [64]. Hu et al. report increased risk taking for drivers in the negative emotional states of Anger and Sadness [50].

### 3.6 Road Element Localization

Previously introduced studies also contribute learnings on situational awareness in emotional situations: road element localization was negatively influenced by both Anger, as reported by Zhang et al. [120], as well as Sadness, shown by Jallais et al. [52]. This indicates that emotion regulation can also be of advantage for the restoration of driver awareness before take-over requests in highly automated vehicles, for example when drivers are emotionally engaged in conversation, gaming, or media consumption.

### 3.7 Subjective Safety Level

A driving simulator study with 70 US undergraduate students by Jeon et al. showed that drivers experiencing Anger reported less subjective safety compared to Neutral and Fear [55]. This can be connected with the increased risk perception of angry drivers reported in Section 3.4. However, it does not reflect the same connection for fearful states.

### 3.8 Workload

Further research by Jeon et al. investigates the impact of emotional states on task workload. They initially assumed higher workload for drivers in Fear from a survey with 14 students [56] but later found no significant differences for load levels of angry, happy, or fearful drivers in a driving simulator setup [55]. Another driving simulator study with 61 undergraduate students suggests that angry drivers are performing worse in self-control, as the author reports higher frustration and more physical workload compared to participants in a neutral state [53]. From these studies, no direct ties between emotional states and workload can be reasoned. We know that cognitive load is connected to an increase in reaction time and an overall impairment of driving performance [30], just like the analyzed work has found for emotional extremes. It would thus be interesting for future work to explore potential correlations between emotions and cognitive load while driving.

### 3.9 Summary

The presented research concentrates heavily on the emotional state of Anger. It shows convincing evidence of Anger being a cause for aggressive driving and worse driving performance, e.g. through an increase of reaction times and an impairment of risk perception. Other negative states such as Fear and Sadness have experienced less attention within the research community but also show negative effects on driving performance and other safety relevant aspects. Similar negative effects on driving performance have also been shown within the few studies incorporating intense Happiness and Contempt.

Though the overview in Table 1 illustrates research gaps regarding specific emotions, the related work confirms the veracity of the simple driver state models by Cai & Lin [21] and Braun & Alt [12], according to which drivers should ideally be in a state of medium arousal and slightly positive valence (see Figures 1 and 2). This can be used as a base to develop user-aware systems which can guide the emotional development of the driver and thus improve road safety.

## 4 TRIGGERS OF EMOTIONS WHILE DRIVING

Additionally to the effects of emotions on the road, a limited amount of research has investigated the context in which pronounced emotional states can develop. The work of Jeon & Walker analyzes online brainstormings with 14 university students to identify driving scenarios that can induce emotions [54]. Braun et al. followed a similar approach with an online study on self-reports from 170 drivers from Germany and the US, addressing triggers for positive and negative emotions while driving [16]. Burns & Katovich examined 152 newspaper articles on road rage and aggressive driving to infer the reasons for such behavior [20]. Underwood et al. had 100 British drivers keeping diaries over two weeks of driving, looking at the factors for Anger in the car [106]. Three studies derive triggers of emotions from observing naturalistic driving: Mesken et al. with 44 Dutch drivers [70], Wurhofer et al. with 10 Austrian and Madeiran drivers [117], and Zepf et al. with 33 German drivers [119]. The identified interrelations are depicted in Table 2, split into negative and positive emotions, as most related work uses this style of classification instead of categorical emotions.

### 4.1 Triggers of Negative Emotions

The most common reasons for negative emotions, such as Anger or Fear, are traffic-related events caused by other drivers [16, 20, 54, 70, 119], environmental influences [16, 20, 70, 117] and the involvement into near accidents [16, 54, 106]. Zepf et al. also report negative emotions triggered through unsatisfactory interaction with the in-car user interface, especially for navigation features [119]. As further sources for negative affect, the driver's personal driving performance [16, 119], verbal interaction with passengers [54], time pressure [20, 54] and insufficient capabilities of the car [16] have also been reported.

### 4.2 Triggers of Positive Emotions

The related work focuses mainly on negative emotional states, as they have been shown to have more severe implications for driving safety. Positive states are usually not elicited through other drivers' behaviors but rather influenced by nice surroundings [16, 119] and personal interactions with passengers [16, 54]. Moreover, drivers gave account of their ability to drive well and their car's performance features to cause positive emotions [16, 119]. Positive emotions consequently are often results of general enjoyment, which is hardly surprising.

## 5 EMOTION REGULATION APPROACHES

Systems which detect and react to the driver's affective state in order to regulate the driver's emotions are thought to have great potential for improving road safety. Emotion regulation is widely used in human interaction, be it for empathizing, e.g. when condoling others, or to influence one's personal emotional state, e.g. to get pumped before an athletic performance [11]. Emotion regulation is broadly defined as "all of the conscious and non-conscious strategies we use to increase, maintain, or decrease one or more components of an emotional response" [40]. The methods to achieve this effect depend on the initial emotion and whether an up- or down-regulation is intended [11]. In automotive contexts we see different approaches for different emotional states, aiming at, e.g., calming angry drivers or energizing fatigued drivers. Experiments on these approaches are laid out in Table 3, for an overview of user-centered research on in-car emotion regulation approaches.

## Triggers of Emotions while Driving

|  | **Negative Emotions** | **Positive Emotions** |
|---|---|---|
| **Driving Behavior, Traffic** | Braun et al. 2018a; Burns & Katovich 2003; Jeon & Walker 2011; Mesken et al. 2007; Zepf et al. 2019 |  |
| **Environment** | Braun et al. 2018a; Burns & Katovich 2003 Mesken et al. 2007; Wurhofer et al. 2015 | Braun et al. 2018a; Zepf et al. 2019 |
| **Interface** | Zepf et al. 2019 |  |
| **Near Accident** | Braun et al. 2018a; Jeon & Walker 2011; Underwood et al. 1999 |  |
| **Own Performance** | Braun et al. 2018a; Zepf et al. 2019 | Braun et al. 2018a |
| **Personal Interaction** | Jeon & Walker 2011 | Braun et al. 2018a; Jeon & Walker 2011 |
| **Time Constraints** | Burns & Katovich 2003; Jeon & Walker 2011 |  |
| **Vehicle Performance** | Braun et al. 2018a | Braun et al. 2018a; Zepf et al. 2019 |

Table 2. Overview of situations found to influence the emotional state of drivers. Most triggers are either accountable for positive or negative influences on driver emotions.

### 5.1 Adaptive Music

Selecting musical pieces according to the driver's emotional state has been proposed manyfold in the literature. Eyben et al., Hernandez et al. and many other researchers envision emotionally adaptive music in the driving context to influence the driver's emotion in a positive way [8, 16, 22, 31, 47, 121]. The first to investigate this application were Pêcher et al. in a static driving simulator study with 17 French drivers [81]. They report that happy music decreased driving performance, while sad music inspired a more calm driving style with better performance. Brodsky & Kitzner also report that calming music increased levels of positive affect from an on-road study with 22 Israeli undergrad students [19]. They further noticed a weak habituation effect, meaning a slight decrease in response to the stimulus after repeated exposition. Another static simulation experiment with 19 Dutch drivers by Van der Zwaag et al. shows that positive and negative music both have a calming effect on respiration compared to no music and in the long run affect the driver state and with that driving behavior [107]. FakhrHosseini et al. have published a series of simulator studies with a total of 132 university students from the US, where participants listening to either happy or sad music made less driving errors than those not listening to music [35] and angry drivers with self-selected music drove more aggressively than without [33]. Yet, angry drivers with self-selected music also reacted faster than without music, comparable to drivers in a neutral state [34]. They also state that although music affected driving performance and reaction times, it had no significant influences on subjective emotion ratings [33].

Another challenge for emotion regulation through music is the recommendation process itself. On one hand, offering adaptive music to influence emotional states is a quite obvious regulation technique which might cause reactance in the user. Recommending calming musical pieces to drivers in high arousal states could, on the other hand, help them to calm down and break the cycle of emotional reinforcement through self-selected music.

The perception of music and accompanying emotions are highly subjective. Thus, we face the fundamental problem that one piece of music can hardly be classified into a category of emotions it will induce. However, previous research shows positive emotional reinforcement occurs with preferred and familiar music, when singing, and when improvising for musical individuals. Conversely, complexity, dissonance, and unexpected sounds generally have a rather negative impact [71]. As a solution to this problem of uncertainty from a researcher's perspective, there emerged approaches to annotate music titles with crowd sourced values for arousal and valence, such as the DEAM Database for Emotional Analysis in Music [2]. Researchers are increasingly relying on such databases instead of subjectively selected music.

### 5.2 Ambient Light

Many modern cars are equipped with light conductors in peripheral surfaces, allowing for ambient illumination of the driver's compartment. Several concepts brought up, e.g., by Eyben et al., Coughlin et al. and others, suggest using this technology to take influence on the driver's emotional state through subliminal light cues [16, 24, 31, 65, 77]. Spiridon & Fairclough have investigated the effects of blue light on anger levels in a static driving simulator with 30 adult participants from the UK [98]. They report decreased subjective feelings of anger and lower systolic blood pressure when the blue light stimulus was active. Hassib et al. give account of improved driving performance with blue and orange ambient light for drivers in negative emotional states, evaluated in a static driving simulation with 12 German adults. However, they ascribe this improvement rather to raised attention than a change of affect, as participants did not show significantly different subjective affect ratings between stimuli [45]. A driving simulator experiment with 60 German drivers by Braun et al. conversely showed decreased lane keeping performance with ambient light stimuli for both angry and sad drivers, though the effects were more substantial within the Anger condition [17].

The findings reported in the articles above suggest that ambient lights can have alarming or distracting qualities for users but they can also have a calming effect, plausibly depending on brightness, position and personal familiarity with the feature.

### 5.3 Empathic Speech

Conversational user interfaces based on natural language understanding and synthesis are steadily becoming ubiquitous for phone assistants, household appliances, and with some delay also for in-car interaction [116]. Empathic adaptation of a voice assistant's speech output to the driver's and passengers' emotional state and to the driving context have naturally been called for by a vast number of related research [12, 16, 31, 41, 47, 91, 108]. Early research on empathic voice interaction by Nass et al. provides evidence that a matching of voice arousal levels can have positive effects on driving performance and on the frequency of interaction with the speech assistant [75]. A study by Hsie et al. with 20 US drivers further brought to light that modeling speech output with Anger can increase alertness levels of drivers and so reduce distracted driving [49]. Braun et al. conducted a comparative study with empathic speech versus ambient light, visual feedback and a neutral voice assistant, where empathic speech was not only most liked by users but also had improving effects on Sadness and Anger [17].

From these finding we can assume that empathic voice interaction is a suitable means to improve driver focus and to empathize with the driver in negative emotional states. As all presented studies worked with prototypical interactions, a clear description for what empathy means in voice interaction is however not present in the current state of the art.

# Affective Interaction to Improve Driving Behavior through Emotion Regulation

| | Anger | Sadness | Frustration | Fatigue | Happiness | Valence | Arousal |
|---|---|---|---|---|---|---|---|
| Adaptive Music | FakhrHosseini et al. 2014<br>FakhrHosseini & Jeon 2016<br>FakhrHosseini & Jeon 2019 | Pêcher et al. 2009 | | | Braun et al. 2018b<br>Pêcher et al. 2009 | Braun et al. 2018b<br>Brodsky & Kitzner 2012<br>Van der Zwaag et al. 2012 | Van der Zwaag et al. 2012 |
| Ambient Light | Braun et al. 2019a<br>Spiridon & Fairclough 2017 | Braun et al. 2019a | | | | Hassib et al. 2019 | Hassib et al. 2019 |
| Empathic Speech | Braun et al. 2019a | Braun et al. 2019a | Nass et al. 2005 | | Nass et al. 2005 | | Hsie et al. 2010 |
| Intervention | Braun et al. 2019a<br>Johnson & McKnight 2009 | Braun et al. 2019a | | | | Braun et al. 2019b | |
| Reappraisal | Lu et al. 2013 | | Harris & Nass 2011 | | | | |
| Relaxation Techniques | | | Balters et al. 2019 | | | | Balters et al. 2018<br>Paredes et al. 2018a<br>Paredes et al. 2018b |
| State- / Bio-Feedback | Braun et al. 2019a<br>Völkel et al. 2018 | Braun et al. 2019a | | Völkel et al. 2018 | | Braun et al. 2019b | |
| Temperature Control | | | | Schmidt et al. 2017<br>Schmidt & Bullinger 2017 | | | |

Table 3. Overview of emotion regulation approaches investigated in the related work, grouped by categorical emotions, valence and arousal.

## 5.4 Interventions

While the methods described above aim to improve the driver state on rather subliminal channels, it is also plausible to blatantly intervene when possibly dangerous affective states are likely to occur. Nasoz et al. initially proposed a simple warning in case negative emotions are detected [74] and later expanded their approach to distracting drivers from the source of negative emotions by, for example, taking a break [73], also proposed by Oehl et al. [77]. The idea of warning drivers of upcoming traffic obstructions to mitigate Anger was investigated by Johnson & McKnight in an on-road study with 55 US drivers [57]. They report a reduction of aggressive driving behavior for participants with high dispositional anger, meaning they are rather likely to get angry, but also an increase in aggressive driving among participants with low dispositional anger, who were precautionarily warned about congestions. Braun et al. implemented a set of visual feedback systems for a study with 328 German participants, among which warnings concerning the current negative emotional state of the driver were assessed as less attractive than a continuous driver state display. Yet, data from a camera-based emotion detection system supported the efficacy of such interventions as they reportedly made drivers cheer up [13]. Distracting the driver from their current state through voice prompts offering entertainment were shown to have equally little merit compared to no interaction in driving simulator experiment with 60 German drivers by Braun et al. [17]: an emotion recognition software recorded an increase in valence but also distraction increased and driving performance went down.

Interventions are distractive by design. Thus, their application while driving needs to be well planned. In addition, uncalled for recommendations can – especially when trying to influence the user – be perceived as paternalizing. The timing of interventions is also crucial for their acceptance, as complex driving scenarios often require too much cognitive work to handle secondary interaction [95].

## 5.5 Reappraisal

A special form of intervention, aimed at illuminating frustrating situations in more positive light, are reappraisals. They have first been investigated in an automotive context by Harris & Nass, who had a voice assistant reframe traffic situations in an either positive or negative way with 36 US university students in a static driving simulator. Drivers who heard reappraisals to deflate the negativity of the situation reported less negative emotions and performed better than with negative or no feedback [43]. Zhang et al. propose reappraisals to tackle higher risk taking for angry drivers [120], which they investigated in a series of lab studies with Lu et al.: a total of 299 Chinese drivers experienced reappraisals along the dimensions of certainty, control and responsibility, resulting in a normalization of risk perception for drivers in angry and fearful states [66]. Reappraisals are tactful interventions which could resolve the potential of paternalism when done right, and possibly nudge the user towards a more positive state and thus safer driving performance.

## 5.6 Relaxation Techniques

An approach to influence the driver state through active behavioral changes are relaxation techniques, such as breathing exercises, as proposed by Nasoz et al. and Oehl et al. [73, 77]. Research by Paredes et al. with 24 US drivers in a static driving simulator shows that guided slow breathing can reduce drivers' breathing rates without affecting driving safety [80]. They investigated instructions through speech output and haptic stimuli, with the latter being rated as less distracting and more natural to follow. Their work was carried on by Balters et al. who implemented fast breathing interventions for drivers in low arousal to increase alertness [7] and slow breathing instructions to calm down aroused drivers [6], which they validated in two studies with a total of 48 US drivers.

Another approach to relaxation in the car by Paredes et al. is utilizing VR to submerge into a calming under water world which adapts to the movements from driving in order to prevent motion sickness [79]. This can, of course, only be experienced by passengers or in fully automated cars, as the VR headset does not allow for driving. The authors report decreased arousal levels for this relaxation technique from an on-road study with 15 US drivers. Active relaxation techniques could become valuable routines to increase well-being during off-times in (semi-) automated vehicles. Usage while driving might however require more research on the driver's cognitive load and the effects of situational awareness during restorative exercises.

### 5.7 State- / Bio-Feedback

Direct feedback on the user's state is well-known from health apps on consumer devices and has been repeatedly suggested for an in-car application [31, 47, 65]. One utilization of this concept was investigated by Völkel et al. in a dynamic driving simulator setup with 70 German participants. They report that users would only want to receive safety-critical information as notifications but generally preferred a continuous status indicator as it is easy to interpret [110]. Similar findings are reported from a static simulation experiment with 328 German participants by Braun et al.: they state that while a continuous display was generally preferred, senior and inexperienced drivers favored a system with less visual elements and sparse notifications [13]. A visual representation of the driver's emotional state as emoji was examined in another study by Braun et al. with 60 German drivers in angry or sad states [17]. Their verdict is that visual mirroring can amplify the user's negative emotional state, which is not accepted by users and needs to be avoided. It seems that direct feedback on the detected driver state has little value for emotion regulation but might rather be a novelty feature.

### 5.8 Temperature Control

The idea of tackling low arousal, especially fatigue, among drivers has been put forward by Eyben et al. and Hernandez et al. [31, 47] and implemented by Schmidt et al. for experiments in a static driving simulator setup [92, 94]. They report from studies with a total of 67 German drivers that cool air streams were connected with a decrease in fatigue and an increase in alertness and sympathetic activity, led to better driving performance and were preferred by the drivers. Temperature control was shown as effective measure against sleepiness and it can be implemented using technology which is already installed in modern cars. Unsurprisingly, a number of manufacturers have already announced related features.

### 5.9 Further Ideas

There exist more ideas on regulating driver emotions in related work which have not yet been investigated enough to be reported here or are somewhat out of scope for this work. Emotionally adaptive GUIs have, for example, been proposed by several researchers [12, 74, 89] and the same goes for emotionally aware navigation concepts [16, 47, 82]. First evaluations of such affective navigation features promise good acceptance and user experience, e.g., for smart routing based on fleet emotion data [14]. Other ideas are a means of communicating the driver state to the outside of the car [47, 113] or sharing emotional states of passengers with the driver to influence their driving [37]. Further research concerns ideas for automated vehicles, allowing for emotion regulation approaches which require the driver's full attention, such as remotely sharing quality time with family and friends [78].

## From Research to Industry

| | **Research Proposals** | **Industry Applications** |
|---|---|---|
| **Adaptive Music** | Bankar et al. 2018; Braun et al. 2018a; Çano et al. 2017 Eyben et al. 2010; Hernandez et al. 2014; Zhu et al.2016 | BMW Caring Car; Mercedes-Benz Energizing Honda Concept NeuV; Toyota Yui |
| **Adaptive UI** | Braun & Alt 2019; Nasoz et al. 2002; Row et al. 2016 | Nio NOMI |
| **Ambient Light** | Braun et al. 2018a, 2019a; Coughlin et al. 2001; Eyben et al. 2010; Löcken et al. 2017; Oehl et al. 2019 | Audi Fit Driver; BMW Caring Car Mercedes-Benz Energizing |
| **Emotion-Aware Navigation** | Braun et al. 2018a; Hernandez et al. 2014 Pfleging et al. 2014 | |
| **Empathic Speech** | Braun et al. 2018a; Braun & Alt 2019; Eyben et al. 2010 Gusikhin et al. 2011; Hernandez et al. 2014 Sarala et al. 2018; Vögel et al. 2018 | Toyota Yui |
| **Intervention** | Braun et al. 2019a; Nasoz et al. 2002, 2010; Oehl et al. 2019 | |
| **Reappraisal** | Braun et al. 2019a; Zhang et al. 2016 | |
| **Relaxation Techniques** | Nasoz et al. 2010; Oehl et al. 2019 | Audi Fit Driver |
| **State- / Bio-Feedback** | Braun et al. 2019a; Eyben et al. 2010 Hernandez et al. 2014; Löcken et al. 2017 | Audi Fit Driver; Ford Buzz Car |
| **Temperature Control** | Braun et al. 2019a; Eyben et al. 2010 Hernandez et al. 2014 | Audi Fit Driver; BMW Caring Car Mercedes-Benz Energizing |

Table 4. Initial ideas from related work and their applications in concept cars and series features.

## 6 INDUSTRY APPLICATIONS

Substantial parts of the reported work were funded or executed by research departments of automotive manufacturers. Thus, it is only natural that some concepts sooner or later find their way into production vehicles. We juxtapose initial research ideas and their applications in the industry, showing that a number of approaches have already found their way to the user in one way or another. Table 4 provides a perspective on ideas from research and how they have been applied in cars up to date. Most notably the premium brands Audi, BMW and Mercedes-Benz have been pushing affective features within their well-being functionalities, although, realistically speaking, most of these features are very limited at the moment. Exemplary for the research endeavors in the automotive sector is the agenda by Vögel et al., in which renowned researchers from universities and research institutes, together with specialists from BMW, have envisioned a possible marching route for emotionally aware in-vehicle assistants [108]. Many other manufacturers also cooperate with universities and start-ups in order to explore new possibilities. Most recently, automaker Karma has introduced an instrumentalized Revero as technology carrier for emotion detection algorithms provided by the start-up Xperi, which they use for experiments on affective interaction [58].

Corporate researchers are often cautious with publishing new concepts in technical literature and rather communicate their ideas in the form of concept cars, likely also because of the increased publicity. The first affective concept of the industry was presented by Ford at the Chicago Auto Show in 2009. Their *Emotive Driver Advisor System (EDAS)* is an emotional speech dialog system with graphical avatar [41]. They also built a car exterior concept called *Buzz Car* which senses the driver's emotional state and reflects it through light animations [36]. Honda's *Concept NeuV* introduces an AI assistant which utilizes an "emotion engine" in combination with knowledge about past behavior to predict choices and offer recommendations in daily driving and music selection [48]. In 2017, Audi debuted the concept car *Elaine*, which includes functionalities to detect stress and fatigue from wearable body temperature and heart rate sensors. These can be used to adapt the car for more relaxing or vitalizing journeys, for example with coaching of relaxing breathing techniques [4]. They recently updated this idea to include the user's overall condition, derived from their driving style and vital functions with the concept *AI:ME* at the 2020 Consumer Electronic Show [5]. Toyota also teases concepts which improve safety through driver state detection [104] and they give a glimpse on what emotional interaction with an AI assistant could look like in a more distant future with their *Concept-i* [103]. Similarly, the KIA *Real-time Emotion Adaptive Driving (R.E.A.D.)* System monitors the driver's emotional state and tailors the interior environment to create a more joyful experience. [59]. In a more artistical thought experiment, Mercedes-Benz introduced the *Vision AVTR* at the 2020 Consumer Electronics Show, which recognizes the user based on their breathing patterns and reacts emotionally on proximity and touch, based on the behavior of creatures within the *Avatar* movie franchise [69].

Apart from concept cars, some manufacturers already offer features aimed at improving the driver's emotional experience. Nio cars come with the digital assistant NOMI, which interacts with the people inside the vehicle in a human-like way, trying to improve the user experience with its cuteness [76]. Mercedes-Benz offers a 10-minute coaching program to energize the driver using climate control, ambient lighting and music to increase alertness on long journeys. The feature can allegedly also incorporate and adapt to physiological data from fitness trackers [68]. Similarly, BMW drivers can experience the *Caring Car* feature, which aims to vitalize drivers when they are fatigued or create a relaxing atmosphere when they are stressed, using interior lighting, roof shading, music adjustments, fragrances, seat ventilation and massaging [10]. Another idea which is already in production with many driving assistant functions is using driver camera systems to assess head postures and gaze in order to allow for longer hands-off times given the driver is sensed as attentive [72]. And in a more safety-oriented application of driver state detection, Volvo plans on using distraction and intoxication sensors to allow the car to intervene if the driver is incapable of driving, in order to prevent traffic accidents [111].

We can see that automotive user interfaces are currently becoming more user-centered, yet truly affective systems are still a thing of the future. While emotion detection technology can already be used with consumer hardware, accuracy levels are not yet in a range for confident interventions while driving. Furthermore, the best modalities and the categorization parameters used for detection are being questioned by many, especially as the effects of emotional states and cognitive load are hard to separate in real-world settings. IT consulting agency Gartner classifies Emotion AI as an emerging technology shortly before its "peak of inflated expectations", which will most likely be used productively in 5–10 years [39]. Considering typical development cycles in the automotive industry, 5 years appear as an appropriate timespan to go from initial research to market readiness. Under this angle, it is reasonable for the manufacturers to hold back with immature product innovations at this moment and rather work on improving natural interaction in the car, which can be enhanced with affective features once the technology is there. Research on affective automotive user interfaces should thus explore concepts for interaction in the near future rather than the present, which will most likely introduce higher levels of automation, and with this new interaction contexts.

## 7 ANALYSIS

The literature at hand provides an overview on the current state of research regarding effects and triggers of emotions on the road, as well as on emotion regulation approaches for automotive contexts. The field seems to have a common understanding of the main effects of emotions and a reasonable model of driver emotions. Most research is focussed on implications of angry driving, which is shown to have strong impacts on road safety. Other emotional states, such as Fear or Happiness, are also investigated and can be positioned within the valence/arousal framework. The fact that driver emotions are so far regarded in isolation, while established research on automotive user interfaces has long accepted distraction and cognitive load as key components for safe interaction while driving, can be seen as an indicator that affective in-car systems are not yet ready for the consumer market. It is also striking that different emotion classifications are approached with different means of regulation, e.g. temperature control for tired drivers or ambient lighting for sad or angry motorists. The goal for upcoming research in the field should be to build upon the prevailing theoretical understanding with hands-on prototypes evaluated under reliable and replicable settings and with more lifelike contextual influences than currently practiced. The following sections aim to give researchers and practitioners an overview of the methods used in the past and what limitations they entail, in order to optimize future experimental design accordingly.

### 7.1 Methods

The presented literature is using diverse methodological approaches we see appropriate to examine more closely. All relevant publications containing user-centered research are categorized in Table 5 according to the study's setting, the source of emotions, the applied measures of driver states and the participant sample. Studies on triggers of emotions have been conducted as surveys and workshops, as well as under real traffic conditions, while experiments on the effects of emotions on driving and emotion regulation approaches were almost exclusively conducted in the driving simulator or, scarcely, in real traffic. This can be explained with the different requirements for these types of studies: the investigation of triggers relies on emotions experienced during real driving, reported through subjective feedback either while driving or in retrospect. Measuring the effects of emotions, however, requires a ground truth of the driver's state, which is most often generated through emotion elicitation techniques. As these studies hypothesize negative effects on driving performance from emotional influences, inducing emotions in real traffic would introduce ethically unacceptable risks for participants, which are circumvented in the driving simulator. Furthermore, quantitative data on driving performance can be easily recorded in a simulator setup. The validity of simulator experiments in comparison to real driving have been shown, e.g., by Klüver et al. [60]. They recommend moving-base simulators for best comparability, yet many studies are conducted in low-fidelity seating bucks with limited immersion and sometimes very artificial driving tasks. This should be considered when assessing the findings which come from such studies.

Studies on emotion regulation techniques also come with a set of limitations readers need to be aware of, for example that artificially induced emotions can fade away after relatively short time. Braun et al. report mean durations of 2–7 ± 2 minutes depending on the used technique, after which participants reported their emotions to have reverted back to their originating state [18]. Thus, it is crucial to include a baseline without emotion induction into tests on emotion regulation techniques. Another peculiarity of interaction with the goal of regulating emotions are possible misapprehensions regarding causalities between state and behavior. Fakhrhosseini and Jeon [33] as well as Hassib et al. [45] both give account of successfully influencing initially angry drivers to drive more safely. Yet, they also report that this was more likely due to increased attention during interactions and not because their emotional state had improved. Finally, as is the case for

## Overview of Methods

| Effects and Triggers of Driver Emotions | N | Survey / Workshop | Laboratory | Driving Simulation | Naturalistic Driving | Emotion Induction | Traffic Scenario | Physiological Data | Driving Performance | Subjective Self-Rating | University Students | Company Employees | General Population |
|---|---|---|---|---|---|---|---|---|---|---|---|---|---|
| | | SETTING | | | | EMO SOURCE | | METRIC | | | SAMPLE | | |
| Braun et al. 2018a | 170 | ● | | | | | | | | ● | ● | | |
| Cai & Lin 2011 | 15 | | ● | | | ● | ● | ● | ● | ● | | | |
| Chan & Singhal 2014 | 25 | | ● | | | ● | | | ● | ● | ● | ● | |
| Deffenbacher et al. 2003 | 121 | | ● | | | ● | | | ● | ● | | | |
| Hu et al. 2012 | 218 | ● | | | | ● | | | ● | | ● | | |
| Hu et al. 2012 | 500 | ● | | | | | | | ● | | ● | | |
| Jallais et al. 2014 | 54 | ● | | | | ● | | | ● | | ● | | |
| Jeon & Walker 2011 | 14 | ● | | | | ● | | | ● | | ● | | |
| Jeon et al. 2011 | 24 | | ● | ● | | ● | | | ● | | ● | | |
| Jeon et al. 2014 | 70 | | ● | ● | | ● | | ● | ● | | ● | | |
| Jeon 2016 | 61 | | ● | ● | | ● | | ● | ● | ● | | | |
| Lee 2010 | 24 | | ● | | | ● | | ● | ● | | | ● | |
| Li et al. 2019 | 24 | | ● | | | ● | ● | | | ● | ● | | |
| Lu et al. 2013 | 299 | ● | | | | ● | | | ● | | ● | | |
| Mesken et al. 2007 | 44 | | | | ● | | ● | ● | ● | ● | | | ● |
| Precht et al. 2016 | 38 | | | ● | | | ● | | | ● | | | ● |
| Roidl et al. 2014 | 79 | | ● | ● | | | | ● | ● | ● | ● | | |
| Steinhauser et al. 2018 | 90 | | ● | ● | | ● | | ● | ● | | ● | | |
| Taubman - Ben-Ari 2011 | 571 | ● | | | | | ● | | | ● | ● | | |
| Trick et al. 2011 | 26 | | ● | ● | | ● | | | ● | ● | ● | | |
| Underwood et al. 1999 | 100 | ● | | | | | | | ● | | | | ● |
| Wurhofer et al. 2015 | 10 | | | | ● | ● | ● | | | ● | | | ● |
| Zepf et al. 2019 | 33 | | | ● | | ● | | ● | | | ● | | |
| Zhang et al. 2016 | 24 | | ● | | | ● | | ● | ● | ● | ● | | |
| Zimasa et al. 2019 | 40 | | ● | ● | | ● | | ● | ● | ● | ● | | |

| Applications of Driver Emotion Regulation | N | Laboratory | Driving Simulation | Naturalistic Driving | Emotion Induction | Traffic Scenario | Physiological Data | Driving Performance | Subjective Self-Rating | University Students | Company Employees | General Population |
|---|---|---|---|---|---|---|---|---|---|---|---|---|
| | | SETTING | | | EMO SOURCE | | METRIC | | | SAMPLE | | |
| Balters et al. 2018 | 8 | ● | | | | ● | ● | | ● | ● | | |
| Balters et al. 2019 | 40 | | ● | | ● | | ● | | ● | ● | | |
| Braun et al. 2018b | 19 | | ● | | ● | | | ● | ● | ● | | |
| Braun et al. 2019a | 60 | | ● | | ● | | ● | ● | ● | ● | | |
| Braun et al. 2019b | 328 | ● | | | | | | | ● | | | ● |
| Brodsky & Kitzner 2012 | 22 | | ● | | ● | | | ● | ● | | | ● |
| FakhrHosseini et al. 2014 | 53 | ● | | | ● | | ● | ● | ● | ● | | |
| FakhrHosseini & Jeon 2016 | 61 | ● | | | ● | | ● | ● | ● | ● | | |
| FakhrHosseini & Jeon 2019 | 52 | ● | | | ● | | ● | ● | ● | ● | | |
| Harris & Nass 2011 | 36 | | ● | | ● | ● | | | ● | ● | | |
| Hassib et al. 2019 | 12 | ● | | | ● | | ● | ● | ● | ● | | |
| Hsieh et al. 2010 | 20 | ● | | | | | | ● | | ● | | |
| Johnson & McKnight 2009 | 55 | | ● | | ● | | ● | | ● | | | ● |
| Nasoz et al. 2010 | 41 | ● | | | ● | | ● | | ● | ● | | |
| Nass et al. 2005 | 40 | | ● | | | | | ● | ● | | | ● |
| Paredes et al. 2018a | 24 | ● | | | ● | | ● | ● | ● | | | ● |
| Paredes et al. 2018b | 15 | | ● | | ● | | | ● | ● | | | ● |
| Pêcher et al. 2009 | 17 | | ● | | ● | | ● | | ● | ● | | |
| Schmidt & Bullinger 2017 | 33 | ● | | | | | | ● | ● | ● | | |
| Schmidt et al. 2017 | 34 | ● | | | | | | ● | ● | ● | | |
| Spiridon & Fairclough 2017 | 30 | ● | | | | | | ● | ● | | | ● |
| Van der Zwaag et al. 2012 | 19 | | ● | | ● | | ● | ● | ● | | | ● |
| Völkel et al. 2018 | 70 | ● | | | ● | | | ● | ● | | | ● |
| Zhu et al.2016 | 30 | ● | | | ● | | ● | ● | | | | ● |

Table 5. Methods used in experimental evaluations of effects and triggers of driver emotions (left) and of applications of driver emotion regulations approaches (right). We distinguish between experiment settings (qualitative, in the lab, simulated or real-world driving), whether emotional states were induced or occured naturally within traffic, and the means of data capture. Furthermore, we take a look at the aspect of sample demographics.

many areas of HCI, we might be experiencing a publication bias towards ignoring non-significant findings. This is on one hand promoted by the funding bodies – who wants to invest more time in an experiment without "noteworthy" results? – but also by the increasing competitiveness of academia. On the other hand it impairs the efficiency of the community, as supposedly promising approaches without available reports of previous studies are sometimes being investigated repeatedly due to a lack of communication.

### 7.2 Participant Samples

Recruiting for interaction experiments requires knowledge about future user groups and the infrastructure to reach these people. The reviewed literature shows that oftentimes the necessary time or funds to do extensive recruiting are lacking, or that studies are designed without a precise concept regarding the desired sample groups. Table 5 shows that over 40% of the analyzed experiments were conducted either with university students or company employees, which does not necessarily mean that the findings are less valid than with nonpartisan participants. It should however call attention to the possibility that, for example, students who are required to partake in an experiment to pass a course might be less motivated to actually perform than a potential user who is curious about personally relevant innovations. Hu et al. describe an experiment in co-location with a DMV office [50], an idea which stands out as positive example because DMV visitors who come for official business regarding their driver's license or car registration are most likely engaged in the current car market and thus might represent a relevant sample group.

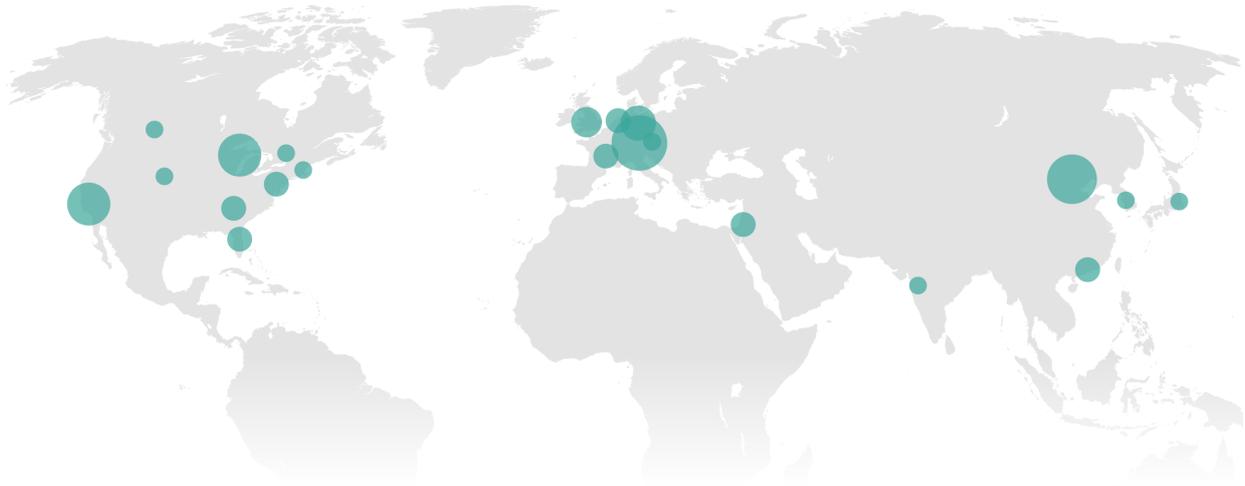

Fig. 3. Research on affective automotive UIs has been predominantly carried out in geographical proximity to major car producers and pertinent research institutes. Circle area represents publication count per region.

Looking at sample sizes, most experiments are executed with a reasonable number of participants. The reviewed surveys are in 3-digit areas, qualitative workshops involve around 10 people and interaction studies, for the most part, report 15–20 participants per between-subject group. The geographical distribution of conducted research, however, clearly shows that affective automotive user interfaces are researched almost exclusively in the vicinity of car manufacturers and at research institutes which specify on in-car interaction. Figure 3 depicts the global distribution of the reviewed studies. The main hubs are located in the Silicon Valley, in the Great Lakes area, in southern Germany and around Beijing, which correlates highly with the locations of development centers of premium vehicles. This is all but surprising, given these three countries make up for roughly half of the annual global car sales [112]. Japanese and Indian manufacturers are underrepresented among the heavyweights, perhaps because, e.g., Toyota communicates more through concept cars than HCI publications (see [103, 104]) and Indian carmakers rather focus on high volume production instead of premium cars with digital services. With these regions as driving forces behind affective technologies, we can expect that future products will be tailored to the needs of the upper and upper-middle class population of highly industrialized countries, ignoring the vast majority of the global population. We cannot foretell whether this is going to cause problems affecting the acceptance of this new technology in the long run but want to highlight the discrepancy at hand.

## 8 DIRECTIONS FOR FUTURE RESEARCH

As concluded in Section 6, consumer market applications of affective technology in the car are likely to emerge within the next 5–10 years. This makes the present the best time to shift from fundamental research to the ideation of features for realistic use cases. The reviewed literature provides unambiguous findings on the impacts of emotions on driving performance and particularly Anger, but also other states of exceedingly high or low arousal have been shown to be of relevance for improving road safety. While the fundamental research on emotions is overall well substantiated, regarding emotional states in isolation strikes far from real circumstances. User experience research in the automotive field has long been incorporating the concepts of distraction and visual, auditory, or cognitive load as some of the main factors for usable in-car interfaces [30, 95]. A universal driver state model encompassing all these reference frameworks is certainly unlikely to evolve in the near future as they embody different views on human behavior. However, a broader perspective for future concepts could be of help.

Another area which might become important for future in-car systems are cultural characteristics of the global user base. We see challenges for emotion detection in disparate behaviors used to express emotions between high and low context cultures (see Hall [42]) and in social acceptance of interacting with affective systems with passengers on board. Lachner et al. propose the design of culture-derived interaction using cultural personas, enabling diverse emotional situations throughout different cultural backgrounds [61]. This could, for example, be used to generate features with users from different regions or demographic backgrounds and iteratively building upon their inputs to establish culturally sensitivity [63]. Future work could investigate cultural adaptation of in-car interaction, for example, with conversational voice assistants. This could also help facilitating a better understanding of driving styles in relation to cultures and emotions, in order to offer perfectly adapted driving assistance.

One viable goal for future systems, apart from deliberate emotion regulation techniques, could also be to improve the user experience, for example through micro interactions specifically designed to evoke positive emotions, improving the atmosphere inside the car. Micro-interactions can, for example, be short interventions in situations where users are idly waiting for a traffic light to turn green [3], or consist of rather decorative graphical or conversational elements to make an interaction more friendly. Hassenzahl proposes the design of well-being-oriented interactive systems, whose designers focus "on creating and shaping enjoyable and meaningful activities through sensible arrangements of interactive technologies" [44]. Such hedonic features are indispensable for the design of good user interfaces in full-attention contexts and might become the gold standard also for automotive user interfaces when systems can justify additional interactions with positive effects on safety-relevant behavior. With mature emotion detection technologies, improvement of hedonic quality in interaction could even be quantified.

Another plausible application area in the near future is driver state control in highly automated vehicles. When drivers still have to take over control from time to time but are generally allowed to divert attention from the driving task and, e.g., take a nap, affective systems could be used to reinstate alertness before a take-over-request is announced. Previous work shows that personalized voice assistants can increase trust in automation [15]. Expanding personalization on an emotional level might be another worthwhile research area contributing to safe (semi-) automated driving.

## 9 CONCLUSION

This review was compiled to give an overview over the current state of affective automotive user interfaces and the rationales behind augmenting cars with emotion regulation technologies. We hope to have aggregated a common understanding of the field for readers from industry and academia. In the first place, the stated goal of affective automotive user interfaces is the (re-)establishment of a safe driver state. Emotion regulation approaches should focus on the down-regulation of high arousal and up-regulation of low arousal, as well as increasing the driver's valence. This is however highly dependent on other influencing factors like the driver's cognitive load and situational context, e.g., behavior of other traffic participants or accompanying passengers. As technology progresses and first applications are being brought to market, we expect improving synergies of personal electronics and vehicle sensors in the interior and exterior, allowing for a more wholistic view on behavioral and contextual determinants.

Affective Automotive User Interfaces                                                                                     137:23